\documentclass[12pt]{article}

\usepackage{amsmath, euscript, amssymb, amsfonts,  latexsym}
\usepackage[matrix,arrow,curve]{xy}

\usepackage{mathrsfs}

\newcommand{\mB}{{\mathscr{B}}}
\newcommand{\mK}{{\mathscr{K}}}

\newcommand{\oR}{{\mathbb R}}
\newcommand{\oH}{{\mathbb H}}
\newcommand{\oC}{{\mathbb C}}
\newcommand{\oZ}{{\mathbb Z}}

\newcommand{\ssb}{{\scriptscriptstyle\bullet}}

\renewcommand{\Re}{\mathop{\mathrm{Re}}\nolimits}
\renewcommand{\Im}{\mathop{\mathrm{Im}}\nolimits}

\newcommand{\grad}{\mathop{\mathrm{grad}}\nolimits}
\newcommand{\eqdef}{\stackrel{\mathrm{def}}{=}}

\begin{document}

\bigskip

\hbox to \textwidth{\hfil FIAN-TD/2017-21 }

\baselineskip=15pt

\vspace{1.5cm}

\begin{center}
{\large\bf Dirac's magnetic monopole and the Kontsevich star product}

\vspace{0.8cm}

{\bf M.~A.~Soloviev}\footnote{E-mail: soloviev@lpi.ru}
\end{center}


 \centerline{\small\sl I.~E.~Tamm Department of Theoretical Physics, P.~N.~Lebedev Physical Institute,}

 \vspace{0.1cm}
 \centerline{\small\sl Russian Academy of Sciences,  Leninsky Prospect~53, 119991 Moscow, Russia}

\vskip 2em

\begin{abstract}
\medskip
We examine relationships between various quantization schemes for  an electrically charged particle in the field of a magnetic monopole. Quantization maps are defined in invariant geometrical terms, appropriate to the case of nontrivial topology, and are constructed for two operator representations.
In the first setting, the quantum operators act on the Hilbert space of sections  of a nontrivial complex line bundle associated with the Hopf bundle, whereas the second approach uses instead a quaternionic Hilbert module of sections of a trivial quaternionic line bundle. We show that these two quantizations are  naturally related by a bundle morphism  and, as a consequence, induce the same phase-space star product. We  obtain explicit expressions for the integral kernels of star-products corresponding to various operator orderings and calculate their asymptotic expansions  up to the third order in the Planck constant $\hbar$.  We also show that the differential form of the magnetic Weyl product corresponding to the symmetric ordering agrees completely with the Kontsevich formula for deformation quantization of Poisson structures and can be represented by Kontsevich's  graphs.

\medskip

\end{abstract}

\medskip

\noindent
PACS numbers: 14.80.Hv, 03.65.Vf, 03.65.Ca, 02.40.Yy,  02.40.Ma


\medskip

\section{\large Introduction}

\baselineskip=20pt

The purpose of this paper is to give a comparative analysis of various quantization schemes
for a charged particle in the presence of a magnetic monopole.
Since the works of \'Sniatycki~\cite{Sn}, Greub and Petry~\cite{GP}, and Wu and Yang~\cite{WY1,WY2,WY3}, it has been generally recognized that the theory of fiber bundles provides the most appropriate  framework for describing the quantum dynamics of this system without using strings of singularities.  This geometric description  reveals the topological origin of Dirac's charge quantization condition~\cite{D1,D2}, shows the role of  the Hopf fibration~\cite{H} in the monopole context, and gives adequate tools to analyse  the symmetry properties~\cite{H1,Ca,H2}. Although there is a vast body of literature on this subject and the charge-monopole system has been deeply studied from various angles (see review by Milton~\cite{Milton}), the geometric and functional analytic  aspects of constructing the Weyl correspondence between symbols and operators in this topologically nontrivial case deserve more study, especially in connection with  developments in the so-called magnetic Weyl calculus~\cite{M,KO1,KO2,MP1,MP2,L}. This calculus extends the usual Weyl symbol calculus~\cite{Fol,W,deG} to magnetic systems, but under the assumption that the phase space topology is trivial and the magnetic field has a globally defined vector potential, which is not the case for the monopole field. Another point deserving attention is that the motion of a quantum particle in the monopole field can also be described by using the quaternionic Hilbert space formulation of quantum mechanics~\cite{A1}. An advantage of this approach proposed by Emch and Jadczyk~\cite{EJ}  is that it deals with a trivial fiber bundle whose sections can be treated as functions. It is also of interest to consider the charge-monopole system from the viewpoint of the theory of deformation quantization of Poisson manifolds (see, e.g.,~\cite{DS} for a review of this area) and to construct the corresponding star-product algebra because this system provides a simple and instructive example of a nonstandard symplectic structure.

The  theory of fiber bundles provides
a  global Lagrangian description of  the charge-monopole system
as a constrained system with $U(1)$ gauge symmetry, see~\cite{BMSS1,BMSS2,S84,P,S16}.
This suggests that this system can be quantized by applying
the usual Weyl quantization map to the phase-space functions lifted to an enlarged phase space which includes the gauge variables and has the standard symplectic structure.
 Such a quantization was studied in~\cite{S16},
   but in the present paper, we construct a quantization map in a different way which is more convenient for computation of the  star product and is closer to the
   definition used in the magnetic Weyl calculus for the  divergence-free  magnetic fields and based on the insertion of a magnetic vector-potential into the Weyl system. The formulation given in Sec.~5 below is an adaptation of this definition to the case when there is no globally defined  vector-potential. It uses an operator representation in the Hilbert space of  sections of a complex line bundle associated with the Hopf bundle and is given in terms of the parallel transport of fibers, which makes it completely gauge independent. Moreover, we construct and study a whole  family of quantization maps
corresponding to various operator orderings,  including,  besides the Weyl ordering, analogs of the standard and anti-standard orderings.

 The question of constructing a Weyl-type map
   in the quaternionic setting proposed  by Emch and Jadczyk was raised by Cari\~{n}ena et al.~\cite{CG-BLMV1,CG-BLMV2}. A
  regular procedure for finding the corresponding
    star product  was developed in~\cite{S}, starting from the multiplier of a quaternionic projective representation  of the translation group, introduced in~\cite{EJ},  and using the Zassenhaus formula for noncommuting operators.  Here we examine its relation to  an alternative approach based on expressing the multiplier in terms of the magnetic flux.

The paper is organized as follows.  Sections~2 and 3 introduce  notation and provide the minimum of information about the Hopf fibration which is necessary for the subsequent analyses. For further details we refer the reader to~\cite{U}; a brief readable sketch of fiber bundle theory can  be found in~\cite{DV}. In sections~4 and 5, we introduce the basic definitions of  magnetic translations and quantization maps formulated in invariant geometrical terms which are appropriate to the case of nontrivial topology. The main new results are presented in sections~6, 7, and 8. First we  show in Sec.~6 that quantizations of
the charge-monopole system with operator representations in complex and quaternionic Hilbert spaces are naturally related by a bundle morphism   which converts the canonical $U(1)$-connection on the Hopf bundle into an $SU(2)$-connection. Using this relation, we
give a simple and rigorous proof of the formula expressing the multiplier of the Emch-Jadczyk representation in terms of the magnetic flux. These results, in turn, are used in Sec.~7 to prove that the operator quantizations with complex and quaternionic Hilbert spaces yield
the same phase-space star product and to obtain explicit expressions for the integral kernels of star-products corresponding to various operator orderings. In Sec.~8, we derive asymptotic expansions of the  products and show that their derivation by means of expanding the magnetic flux entering in the expression for the multiplier is equivalent to the derivation by using the Zassenhaus formula.  We also show that the differential form of the magnetic Weyl product corresponding to the symmetric ordering agrees completely with the Kontsevich formula~\cite{K} for deformation quantization of Poisson manifolds. The asymptotic  expansion of this product  is explicitly expressed,  up to the third order in the Planck constant $\hbar$, in terms of the initial Poisson structure   and is represented by Kontsevich's graphs with the identification of the relevant graphs.
  Section~9 contains concluding remarks. Some technical details regarding the calculation of the magnetic star product with the use of the Zassenhaus formula are  given in Appendix.

\section{\large Magnetic Poisson brackets}

It is well known (see, e.g.,~\cite{NT}, Sec.~13.1) that the   equations of motion for a charged particle in a magnetic field $\mathbf B(x)$ can be written in Hamilton's form
\begin{equation}
\dot x^i= \{x^i,H\},\qquad \dot p_i=\{p_i, H\}
 \notag
\end{equation}
taking the kinetic energy as Hamiltonian, that is, setting $H=\dfrac{1}{2m}\sum p^2_i$, and
using the magnetic Poisson brackets
\begin{equation}
\{x^i, x^j\}=0,\quad \{x^i, p_j\}=\delta^i_j,\quad  \{p_i, p_j\}=\beta_{ij}(x),\quad \text{where  $\beta_{ij}=e\epsilon_{ijk} B^k$},
 \label{2.2}
\end{equation}
or, equivalently, using the  symplectic form
\begin{equation}
dp_i\wedge dx^i+\frac12 \beta_{ij} \,dx^i \wedge dx^j
 \label{2.3}
\end{equation}
which corresponds to the Poisson matrix
\begin{equation}
\mathcal P=\begin{pmatrix}0 &I\\-I& \beta(x)\end{pmatrix}.
\label{2.4}
\end{equation}
If the magnetic field has a globally defined vector potential, i.e., $\beta_{ij}=e(\partial_iA_j-\partial_jA_i)$,  then the symplectic form~\eqref{2.3} can be put into canonical form by  changing  variables from $p_i$ to $p_i+eA_i$, but this is impossible for the monopole field
\begin{equation}
 B^k(x) =\mathit g\frac{x^k}{|x|^3},
 \label{2.5}
\end{equation}
where $\mathit g$  is the monopole strength. For this field, it is usual to use two local vector potentials  expressed in spherical coordinates by
\begin{equation}
 \mathbf A_+(r,\phi,\theta) = \frac{\mathit g}{r}\tan\frac{\theta}{2}\,\mathbf
  e_\phi,\,\,\theta\ne\pi,\qquad  \mathbf A_-(r,\phi,\theta) = -\frac{\mathit g}{r}\cot\frac{\theta}{2}\,\mathbf e_\phi,\,\,\theta\ne0,
 \label{2.6}
\end{equation}
(where $\phi$ and $\theta$ are the azimuthal and polar angles, respectively) and related by a gauge transformation in their common domain of definition,
\begin{equation}
 \mathbf A_+=\mathbf A_- +2\mathit g\grad\phi, \quad \theta\ne0,\pi.
 \label{2.7}
\end{equation}
 The explicit form of $\mathbf A_\pm$ is really irrelevant for the basic definitions given below in a gauge-invariant manner, but the very existence of continuously differentiable vector potentials in regions  covering the configuration space $\dot\oR^3=\{x\in\oR^3: x\ne0\}$ is used in proving  some intermediate  statements. The origin is excluded from the configuration space because it is a  point of singularity; in other words, the charged particle and the monopole  cannot occupy the same point in space at the same moment.

\section{\large The  fiber bundle description}

The Schr\"odinger equation dictates that the gauge transformation~\eqref{2.7}  is accompanied by the corresponding transformation of the particle's wave function
\begin{equation}
 \Psi_+=\Psi_-\exp{\left(i\frac{2e\mathit g}{\hbar}\phi\right)},
 \label{3.1}
\end{equation}
and the requirement of consistency between the two local descriptions leads to the charge quantization condition
\begin{equation}
e\mathit g=\frac{1}{2}n\hbar,\qquad n\in\oZ.
 \label{3.2}
\end{equation}
Accordingly,  the kinetic momentum operator $P$ has the  two local representations
\begin{equation}
 P_j=-i\hbar\partial_j- eA_{(\pm)j}.
 \label{3.3}
\end{equation}
Under the condition~\eqref{3.2}, the pair of functions $\Psi_+$, $\Psi_-$ satisfying~\eqref{3.1}  on the overlap of their domains can be treated geometrically as a section $\Psi$ of a complex line bundle $E_n$ determined by the transition function $\exp{(in\phi)}$, and the operator $\partial_j-i(e/\hbar) A_{(\pm)j}$ defines a covariant differentiation in this bundle. Since $\overline{\Phi}_+\Psi_+=\overline{\Phi}_-\Psi_-$ for $\theta\ne 0,\pi$, the scalar product of two sections is naturally defined by
\begin{equation}
 \langle \Phi,\Psi\rangle=\int_{\dot\oR^3}\overline{\Phi}_\pm(x)\Psi_\pm(x) dx.
 \label{3.4}
\end{equation}
Letting  $\mathcal H_n$ denote the Hilbert space of sections equipped with this scalar product and  $Q$ denote the position operator defined by $(Q^i\Psi_\pm)(x)=x^i\Psi_\pm(x)$, we have the commutation relations
\begin{equation}
[Q^i,Q^j]=0,\quad [Q^i,P_j]=i\hbar\,\delta^i_j,\quad\text{and}\quad [P_i,P_j]=i\hbar\, \beta_{ij},
 \label{3.5}
\end{equation}
corresponding to the magnetic Poisson brackets~\eqref{2.2}.

The outlined local (and gauge-dependent) description going back to Wu and Yang~\cite{WY3} is quite sufficient in most instances, but in order to properly define the magnetic translation operator $e^{iu\cdot P}$ and to construct a Weyl-type quantization map in a rigorous and invariant manner, we need some more  concepts from  fibre bundle theory. First, it is useful to consider the line bundle $E_n$ as associated with a principal bundle.
The total space $\dot\oC^2=\{z=(z_1,z_2)\in\oC^2: z\ne0\}$ of the underlying principal bundle is merely an enlarged configuration space which includes the gauge variable and  can be parameterized by two complex numbers~\cite{T}.  Its projection $\pi$ to the initial configuration space $\dot\oR^3$ is defined by
\begin{equation}
\dot\oC^2\stackrel{\pi}{\longrightarrow}\dot\oR^3 \colon \quad x^j=z^\dagger\sigma_j z, \quad j=1,2,3,
 \label{3.6}
\end{equation}
 where $\sigma_j$ are the Pauli matrices.
The gauge group $U(1)$ acts freely on the punctured space $\dot\oC^2$:
\begin{equation}
\dot\oC^2\times U(1)\ni (z, e^{i\alpha}) \longrightarrow z e^{i\alpha}\in \dot\oC^2,
 \notag
\end{equation}
and $\pi(z)=\pi(z')$  if and only if $z'=ze^{i\alpha}$ for some  $\alpha$, i.e., $\dot\oR^3=\dot\oC^2/U(1)$. The restriction of the principal bundle $(\dot\oC^2,\dot\oR^3,\pi,U(1))$ to the unit sphere $S^3$
is just the Hopf bundle
\begin{equation}
S^3\approx SU(2)\longrightarrow SU(2)/U(1)\approx S^2,
 \notag
\end{equation}
where $S^3$  is identified with the group $SU(2)$ by
\begin{equation}
z=(z_1,z_2)\longrightarrow\begin{pmatrix}
z_1&-\bar z_2\\z_2& \bar z_1\end{pmatrix},\qquad |z|^2=1,
 \notag
\end{equation}
The canonical connection on $(\dot\oC^2,\dot\oR^3,\pi,U(1))$,  corresponding to that on the Hopf bundle, is globally  given  by
\begin{equation}
\omega=i\Im(z^\dagger d z)/z^\dagger z.
 \label{3.7}
 \end{equation}
It determines covariant  differentiation in  associated vector bundles and  the parallel transport of vectors along  curves in the base space $\dot\oR^3$ as is explained below.
The complex line bundle  $E_n$ is associated with this principal
bundle   by the  representation
\begin{equation}
U(1)\times\oC\ni (e^{i\alpha}, \zeta)\longrightarrow e^{-in\alpha}\zeta\in\oC.
\label{3.8}
\end{equation}
of the structure group $U(1)$ on the complex plane $\oC$.
   We refer the reader to~\cite{KN} for a precise definition of an  associated vector bundle. An important point is that each $s\in \pi^{-1}(x)$  defines a one-to-one mapping of the standard fiber  $\oC$ onto the fiber of $E_n$
     over the point $x$ and this mapping has an  equivariance property with respect to the action of the structure group on the principal bundle space and on the standard fiber.
   The image of an element of the fiber of  $E_n$ under this mapping can be called its coordinate with respect to $s$. Fixing a local section $s(x)$ of the principal bundle $(\dot\oC^2,\dot\oR^3,\pi,U(1))$ over a $U\subset  \dot\oR^3 $, we  can locally  represent sections of $E_n$ by complex valued functions and  the pullback $s^*\omega$   of the connection form~\eqref{3.7}   yields a local potential. In this way we reproduce   the Wu-Yang description.
In particular,  using the local sections of $(\dot\oC^2,\dot\oR^3,\pi,U(1))$ given by
\begin{equation}
s_+\colon\, z_1=\sqrt{r}\cos(\theta/2),\, z_2=\sqrt{r}\sin(\theta/2)e^{i\phi}
\qquad (\theta\ne\pi)
\label{3.9}
\end{equation}
and
\begin{equation}
 s_-\colon\, z_1=\sqrt{r}\cos(\theta/2)e^{-i\phi},\, z_2=\sqrt{r}\sin(\theta/2)\qquad (\theta\ne 0),
 \label{3.10}
 \end{equation}
we obtain
\begin{equation}
n\,s^*_{(\pm)}\omega=i\frac{e}{\hbar} A_{(\pm)j}dx^j
 \notag
 \end{equation}
with $\mathbf A^{(\pm)}$ defined by~\eqref{2.6}.

Let $\tau=x_t$, $0\le t\le1$  be a path in $\dot\oR^3$ starting from $x$ and ending at $y$  and let $s$ be a local section of $(\dot\oC^2,\dot\oR^3,\pi,U(1))$ over an open set  $U$ containing this path.  Every element of $\pi^{-1}(x)$ can be uniquely written as $s(x)\mathrm g$ with $\mathrm g\in U(1)$. The parallel transport of this element along the path $\tau$ is expressed by
\begin{equation}
s(x)\mathrm g\longrightarrow \exp\left\{-\int_\tau s^*\omega\right\}s(y)\mathrm g.
 \notag
 \end{equation}
Let $s'$ be another section over $U$. Then $s'=s\gamma$ with some $\gamma(x)$ taking values in $U(1)$ and ${s'}^*\omega=s^*\omega+\gamma^{-1}d\gamma$. It follow that $s'$ yields another expression for the same mapping $\pi^{-1}(x)\to\pi^{-1}(y)$ depending only on $\omega$ and $\tau$.
If $\psi$ is an element of the fiber of $E_n$ over $x$ and $\zeta\in \oC$ is its coordinate with respect to $s$, then by definition~\cite{KN}, the element parallel transported from $\psi$ along $\tau$ has the coordinate $\exp\left\{n\int_\tau s^*\omega\right\}\zeta$
with respect to $s(y)$ and does not  dependent on the choice of $s$.

\section{\large Magnetic  translations}

Using the connection form~\eqref{3.7} and the  parallel transport  of fibers we can define a unitary group generated by the covariant derivative in a fixed direction. Namely, let $a$ be a vector in $\oR^3$,
and let $V(a)$ be the  operator  that transforms any section $\Psi$ of the line bundle $E_n$ into another section whose value at the point $x$ is the parallel transport of  the value of $\Psi$ at the point $x+a$ along the straight line path connecting these points
\begin{equation}
x_t=x+a-ta,\quad 0\le t\le 1.
\notag
\end{equation}
There is a subtlety here because this path is contained in the base space $\dot\oR^3$ only if it does not intersect the origin. For this reason, the transformed section is not defined for all $x$ in the closed interval
from the origin to $a$. But this set  has zero Lebesgue measure and the section is therefore well defined as an element of the Hilbert   space $\mathcal H_n$ of square integrable sections. A local expression for $V(a)$ is given by
\begin{equation}
\big(V(a)\Psi_s\big)(x)=\exp\left\{n
\int_{[x+a,x]}s^*\omega \right\} \Psi_s(x+a),
\notag
\end{equation}
where $s$ is a local section of $(\dot\oC^2,\dot\oR^3,\pi,U(1))$ over an open set containing the path $x_t$ and $\Psi_s$ is the complex-valued function which locally represents $\Psi$ with respect to~$s$. Thus, $V(a)$ is  well defined as a unitary operator acting in the space of sections. This is in distinction to the usual magnetic Weyl calculus, where the Weyl system is formed by operators acting in the space of complex-valued functions and the systems corresponding to different choices of the magnetic vector potential are connected by a unitary transformation.

 The product of operators $V(a)$, $V(b)$, and  $V(a+b)^{-1}$ performs the parallel transport of $\Psi(x)$ along the loop  forming the  boundary of the plane triangle $\bigtriangleup(x;a,b)$ with vertices $x$, $x+a$, and $x+a+b$.
Hence, the composite operator merely multiplies $\Psi(x)$ by an exponential phase factor which is determined by the corresponding element of the holonomy group of the connection~\eqref{3.7} with reference point $x$ and also by the representation~\eqref{3.8}. This phase can be written in terms of the circulation of a local vector potential around the triangular loop:
\begin{multline}
\Big(V(a)V(b)V^{-1}(a+b)\Psi\Big)(x)=\exp\left\{-n
\oint_{\partial\bigtriangleup(x;a,b)}s^*\omega \right\} \Psi(x)\\=\exp\left\{-\frac{ie}{\hbar}
\oint_{\partial\bigtriangleup(x;a,b)}\mathbf A\cdot d\mathbf r\right\}\Psi(x)
\label{4.1}
\end{multline}
(where the orientation of $\partial\bigtriangleup(x;a,b)$ corresponds to the sequence $x\to x+a\to x+a+b$). But it is  independent on the choice of $\mathbf A$ and is expressed, by Stokes' theorem, as the  flux of the monopole   field~\eqref{2.5}  through the triangle $\bigtriangleup(x;a,b)$, or equivalently, as the  surface integral of the  magnetic symplectic form $\beta=\sum_{i<j} \beta_{ij}dx^i\wedge dx^j$ over this triangle. Indeed, assuming that  $\bigtriangleup(x;a,b)$ lies inside the domain of regularity of the potential, we have
\begin{multline}
e\oint\limits_{\partial\bigtriangleup(x;a,b)}\mathbf A\cdot d\mathbf r=e\int\limits_{\bigtriangleup(x;a,b)}\mathbf B\cdot d\mathbf s=\int\limits_{\bigtriangleup(x;a,b)}\beta\\=\int_0^1dt_1\int_0^{t_1}dt_2\, a^i\beta_{ij}(x+t_1 a+t_2 b)b^j,
\label{4.2}
\end{multline}
where the natural parametrization $(t_1,t_2)\to x+t_1 a+t_2 b$  is used in the last equality.

Letting $M(a,b)$ denote the operator of multiplication by  the function
\begin{equation}
m(x;a,b)=\exp\left\{-\frac{i}{\hbar}\int_{\bigtriangleup(x;a,b)}\beta\right\},
\label{4.3}
\end{equation}
we may rewrite~\eqref{4.1} as
 \begin{equation}
V(a)V(b)=M(a,b)V(a+b).
\label{4.4}
\end{equation}
Thus, the mapping  $a\to V(a)$ can be thought of as a generalized projective representation of the translation group, for which the exponential of the magnetic flux plays the role of a multiplier.
However, in contrast to the usual projective representations, this multiplier  is   not a scalar function, because it depends nontrivially on the position variable $x$. In particular, $M(a,b)$  does not commute with $V(a+b)$ and the left multiplier should be distinguished from the right one. The associativity of the operator product $(V(a)V(b))V(c)=V(a)(V(b)V(c))$ implies that   $M(a,b)$ satisfies the  2-cocycle  condition
\begin{equation}
M(a,b)M(a+b,c)=V(a)M(b,c)V(a)^{-1}M(a,b+c),
\label{4.5}
\end{equation}
where $V(a)$ cannot be dropped because of the noncommutativity and $V(a)M(b,c)V(a)^{-1}$ is the operator of multiplication by
\begin{equation}
\exp\left\{-\frac{i}{\hbar}\int_{\bigtriangleup(x+a;b,c)}\beta\right\}.
\notag
\end{equation}
In terms of the magnetic flux, the  identity~\eqref{4.5} is  interpreted as stating that the
flux through the surface of the tetrahedron spanned by the points x, x + a, x +
a + b,  and x + a + b + c is an integer multiple of $2\pi\hbar/e$, which is automatically
satisfied by the charge quantization condition~\eqref{3.2}. The interrelation between the associativity condition and the charge quantization has been elucidated by Jackiw and the operators $V(a)$ defined via parallel transport in $E_n$ represent a rigorous realization of the finite translations considered in~\cite{J1,J2}.

\section{\large The magnetic Weyl transform  and other quantization maps}

For each fixed $a\in\oR^3$, the operator-valued function $V(ta)$, $t\in \oR$, is a strongly continuous one-parameter unitary group with infinitesimal generator $-i\nabla_a$, where  $\nabla_a=a\!\cdot\! \nabla$ is the covariant derivative in the direction of $a$.
Now we can define a Weyl-type quantization map for the charge-monopole system  by
taking  the Weyl system to be
\begin{gather}
T(u,v)= V(\hbar u)e^{i v\cdot  Q}e^{-i\hbar  u\cdot v/2}=e^{i(u\cdot  P+v\cdot  Q)}, \quad P=-i\hbar\nabla.
 \label{5.1}
\end{gather}
This system  forms a weak projective representation of  phase-space translations. Indeed, using~\eqref{4.4} and the commutation relation
 \begin{equation}
 e^{i v\cdot Q}V(\hbar u')=e^{-i\hbar u'\cdot v}V(\hbar u')e^{i v\cdot Q}
\notag
\end{equation}
and     letting $w$ denote for brevity the pair of variables $(u,v)$, we find that
\begin{equation}
T(w)T(w')=\mathcal M_\hbar(Q;w,w')T(w+w'),
\label{5.2}
 \end{equation}
where  $\mathcal M_\hbar(Q;w,w')$ is the operator of multiplication  by
 \begin{equation}
\mathcal M_\hbar(x;w,w')= \exp\left\{\frac{i\hbar}{2}(u\cdot v'- v\cdot u')\right\}m(x,\hbar u, \hbar u'),
 \label{5.3}
\end{equation}
with $m(x,\cdot,\cdot)$  given by~\eqref{4.3}. The  multiplier $\mathcal M_\hbar(Q;w,w')$ is  thereby expressed in purely symplectic terms.
The quantization map can now be defined in complete analogy with the usual Weyl correspondence, by substituting the momentum and position operators $P$ and $Q$ into the Fourier expansion of phase-space functions, namely,
 \begin{equation}
f \longmapsto  \mathcal O(f)=\frac{1}{(2\pi)^3}\int\! du dv\,\tilde f(u,v)\,e^{i(u\cdot  P+v\cdot  Q)},
 \label{5.4}
\end{equation}
where
 \begin{equation}
\tilde f(u,v) =\frac{1}{(2\pi)^3}\int\! dxdp\, f(x,p)\, e^{-i(u\cdot p+v\cdot x)}.
 \notag
\end{equation}
This map is clearly well-defined for all functions whose Fourier transforms are integrable.
Following the standard terminology, we say that $f$ is the magnetic Weyl symbol of the operator $\mathcal O(f)$.

It is well known that the simplest way to get operator orderings different from the fully symmetric Weyl ordering is to insert a phase factor of the form $e^{i\hbar tu\cdot v}$ into the Weyl system. A  generalization is achieved by replacing the real parameter $t$ with a matrix
of real numbers, and the corresponding star  product algebras were considered, e.g.,
in~\cite{BD,S13,S14} for the case of a linear phase space
and with the emphasis  on the functional analytic aspects in the last two papers.
Below we extend some of this results to the charge-monopole system. Let $\alpha$ be a $3\times 3$ real matrix and $T_\alpha(u,v)$ be defined by
\begin{equation}
T_\alpha(u,v)= V(\hbar u)e^{i v\cdot  Q}e^{i\hbar v\cdot(\alpha-I)u}=T(u,v)e^{i\hbar ( v\cdot\alpha u-v\cdot u/2)},
 \label{5.5}
\end{equation}
where $ v\cdot\alpha u=v^i\alpha_{ij}u^j$. Then~\eqref{5.2} changes to
 \begin{equation}
T_\alpha(w)T_\alpha(w')=\mathcal M_{\alpha,\hbar}(Q;w,w')T_\alpha(w+w'),
\label{5.6}
 \end{equation}
where  $w=(u,v)$ as before, and
 \begin{equation}
\mathcal M_{\alpha,\hbar}(x;w,w')= \exp\left\{i\hbar (v'\cdot(I-\alpha) u- v\cdot \alpha u')\right\}m(x,\hbar u, \hbar u').
 \label{5.7}
\end{equation}
We will consider the family of quantization maps
 \begin{equation}
f \longmapsto  \mathcal O_\alpha(f)=\frac{1}{(2\pi)^3}\int\! du dv\,\tilde f(u,v)\,T_\alpha(u,v),
 \label{5.8}
\end{equation}
which clearly contains the Weyl transformation~\eqref{5.4} as a particular case specified by $\alpha=\frac12 I$. If $\alpha=0$, then $T_\alpha=e^{i v\cdot  Q} e^{i u\cdot  P}$ and the map~\eqref{5.8} takes each monomial in the variables $x^i$ and $p_j$ into an operator monomial  with the operators $P_j$ placed to the right of all $Q^i$, i.e., we have an analog of  the standard ordering.
Setting $\alpha=I$ gives an analog of the anti-standard ordering, but it should be borne in mind that in both cases the ordering of the operators $P_j$ is left symmetric.

\section{\large Quantization with the use of a quaternionic Hilbert space}

The quaternionic quantization scheme is applicable to the charge-monopole system  because the Hopf bundle is  obtained from a trivial $SU(2)$-bundle by reducing the structure group $SU(2)$ to the subgroup $U(1)$.  Correspondingly, the principal bundle $(\dot\oC^2,\dot\oR^3,\pi, U(1))$ defined in Sec.~3, being homotopically equivalent to the Hopf bundle,
is also obtained from a trivial principal  bundle in a similar manner. This means that there is a bundle morphism described by the commutative  diagram
\begin{equation}
\xymatrix@=2cm{
 \dot\oC^2\ar[r]^(0.35)h\ar[d]_\pi&\dot\oR^3\times SU(2)\ar[dl]^{\rm pr_1}\\
 \dot\oR^3
}
\notag
\end{equation}
where $SU(2)$ acts by right multiplication and the  map $h$ agrees with the group action, i.e.,
$h(z e^{i\alpha})=h(z)\eta(e^{i\alpha})$      with $\eta$ denoting the natural inclusion of  $U(1)$ into $SU(2)$. The map $h$ takes each  $z$ to the pair $(\pi(z), \gamma(z))$, where $\pi(z)$ is given by~\eqref{3.6} and $\gamma(z)$ is defined by
\begin{equation}
  \gamma(z)=\frac{1}{|z|}\begin{pmatrix}z_1&-\bar z_2\\
 z_2&\bar z_1\end{pmatrix}\in SU(2).
   \notag
\end{equation}
Now let  $\Omega$ be the image of the connection~\eqref{3.7} under $h$ and let $\xi$ be a vector field on $\dot\oC^2$. By general results on bundle morphisms (see, e.g.,~\cite{KN}, Chapter~II, Proposition~6.1), we have
\begin{equation}
 (h^*\Omega)(\xi)=\eta_*(\omega(\xi)),
   \notag
\end{equation}
where $h^*\Omega$ is the pullback of $\Omega$ under $h$ and $\eta_*$ is the  homomorphism of Lie algebras induced by $\eta$,
\begin{equation}
 \eta_*\colon \Im\oC\to \mathfrak{su}(2),\quad \eta_*(i)=i\sigma_3.
   \notag
\end{equation}
If we use local sections of the trivial principal bundle $\dot\oR^3\times SU(2)$ corresponding to given local sections of $(\dot\oC^2,\dot\oR^3,  \pi, U(1))$, then the local expressions for $\Omega$  are almost the same  as
 those  for  $\omega$  with only difference that  the imaginary unit $i$ is  replaced by $i$ times the third Pauli matrix. In particular,  using $s_+$  defined by~\eqref{3.9}, we have
\begin{equation}
 s_+^*\omega=\frac{i}{2}(1-\cos\theta)d\phi\quad \text{and}\quad (h\circ s_+)^*\Omega=\frac{i}{2}\sigma_3(1-\cos\theta)d\phi.
   \label{6.0}
\end{equation}
Let $s$ be the canonical global section of the trivial $SU(2)$-bundle defined by $s(x)=(x,e)$. Clearly,
\begin{equation}
 s=(h\circ s_+)\mathrm g,\quad\text{where}\quad \mathrm g(\theta,\phi)= \begin{pmatrix}\cos(\theta/2)& \sin(\theta/2)e^{-i\phi}\\
 -\sin(\theta/2)e^{i\phi}&\cos(\theta/2)\end{pmatrix}.
   \label{6.0*}
\end{equation}
Therefore, by the gauge transformation formula,
\begin{equation}
 s^*\Omega=\mathrm g^{-1}(h\circ s_+)^*\Omega\,\mathrm g +\mathrm g^{-1}d\mathrm g
   \notag
\end{equation}
and a simple computation gives
 \begin{equation}
 s^*\Omega=-\frac{i}{2}\epsilon_{ijk}\frac{x^i}{|x|^2}\sigma^jdx^k.
   \label{6.1}
\end{equation}
The gauge-transformed $\mathfrak{su}(2)$-potential~\eqref{6.1} is  regular everywhere in the base space $\dot\oR^3$.

We let $\mathsf E$ denote the    quaternionic line bundle associated with the principal bundle $\dot\oR^3\times SU(2)$  by identifying  $SU(2)$ with the group
of unit quaternions  and its Lie algebra with the space of  imaginary quaternions.
In particular, the basic quaternionic imaginary units are identified with the Pauli matrices multiplied by $-i$,
\begin{equation}
\mathbf e_j=-i\sigma_j,\quad j=1,2,3.
   \notag
\end{equation}
As usual, we assume that the
 unit quaternion group acts  on  the algebra   $\oH$ of quaternions (i.e., on the typical fiber of $\mathsf E$) by left multiplication.
The  covariant derivative defined by the connection  $\Omega$ on   $\mathsf E$ is written as
\begin{equation}
 \boldsymbol{\nabla}_k=\partial_k+\frac12\epsilon_{ijk}\frac{x^i}{|x|^2}\mathbf e_j
   \label{6.2}
\end{equation}
and  its components  satisfy the commutation relations
\begin{equation}
 [\boldsymbol{\nabla}_i,\boldsymbol{\nabla}_j]=-\frac12 \epsilon_{ijk}\frac{x^k}{|x|^3}\mathbf j(x),\quad\text{where}\quad \mathbf j(x)=\frac{x^k\mathbf e_k}{|x|}.
   \label{6.3}
\end{equation}
The formulas~\eqref{6.2} and~\eqref{6.3} were used in~\cite{EJ} as a starting point.
Clearly, $\mathbf j(x)^2=-1$ and the commutation relations~\eqref{6.3} correspond to the initial Poisson brackets for  the monopole of lowest strength, but
with the  imaginary unit quaternion $\mathbf j(x)$ instead of the complex imaginary unit $i$ occurring in the  case of the $U(1)$ covariant derivative $\nabla$. Let $L^2(\oR^3,\mathbb H)$ be the space of square integrable sections of $\mathsf E$ identified with  quaternion-valued functions on $\oR^3$.
 Following~\cite{A1,EJ}, we assume that $L^2(\oR^3,\mathbb H)$  is a right module over  $\mathbb H$, i.e., the multiplication of its elements by quaternionic scalars is taken to act from the right, while linear  operators act from the left. The inner product on this space is defined by
\begin{equation}
 \langle\mathsf \Phi,\mathsf \Psi\rangle=\int_{\oR^3}\!\overline{\mathsf \Phi}(x)\mathsf\Psi(x) dx,
 \notag
\end{equation}
where the bar denotes the quaternionic conjugation. It is easily seen that he operator of left multiplication by $\mathbf j(x)$ is  unitary and anti-Hermitian and commutes with
 all $\boldsymbol{\nabla}_i$.

The finite translation operators $\mathsf V(a)$  generated by $a\cdot\boldsymbol{\nabla}$, $a\in\oR^3$, can be constructed in a manner analogous to that for the complex line bundle with the minor  simplification that $\Omega$ has a globally defined potential. Namely, we define
\begin{equation}
\big(\mathsf V(a)\mathsf \Psi\big)(x)= \exp\left\{-\int_{[x+a,a]}s^*\Omega\right\}\mathsf \Psi(x+a)
   \label{6.4}
\end{equation}
The operators $\mathsf V(a)$ form a quaternionic weak projective representation of the translation group,
\begin{equation}
\mathsf V(a)\mathsf V(b)=\mathsf M(a,b)\mathsf V(a+b),\quad \big(\mathsf M(a,b)\mathsf \Psi\big)(x)=\mathrm P\exp\left\{\int_{\partial\bigtriangleup(x;a,b)}s^*\Omega\right\}
\mathsf\Psi(x).
 \label{6.5}
 \end{equation}
The multiplier $\mathsf M(a,b)$ in~\eqref{6.5} contains a path-ordering operator $\mathrm P$ because the $\mathfrak{su}(2)$-potential~\eqref{6.1} is non-Abelian. It is worth noting at this point that  the concept of a weak projective representation has been proposed and analysed  by  Adler~\cite{A1,A2}  just in the context of quaternionic Hilbert space.
From~\eqref{6.1}, it is easy to get an explicit expression for the quaternionic phase factor in~\eqref{6.4}, see~\cite{S} for details of this calculation. The multiplier $\mathsf M(a,b)$ was expressed in~\cite{S} by a series expansion obtained by applying the Zassenhaus formula. Another useful expression can readily be obtained from the above-discussed relation between the connection~\eqref{3.7}
and the $\mathfrak{su}(2)$-potential~\eqref{6.1}.

{\bf Theorem~1.} {\it In terms of the magnetic flux, the multiplier $\mathsf M(a,b)$ in~\eqref{6.5}  is expressed as the operator of multiplication by
\begin{equation}
\mathsf m(x,a,b)=\exp\left\{-\frac{\mathbf j(x)}{\hbar}\int_{\bigtriangleup(x;a,b)}\beta\right\},
 \label{6.6}
\end{equation}
where $\beta$ is the magnetic symplectic form for $n=1$.

Proof.} We can use the  local section $h\circ s_+$ for computing  elements of the holonomy group of $\Omega$.  Then the path-ordered exponential reduces to the ordinary exponential and from~\eqref{6.0} we  find immediately that
the holonomy group element determined by the triangular loop $\partial\bigtriangleup(x;a,b)$ at the reference point
$(h\circ s_+)(x)$ is given by
\begin{equation}
\exp\left\{\frac{i\sigma_3}{\hbar}\int_{\bigtriangleup(x;a,b)}\beta\right\}.
\label{6.7}
\end{equation}
 According to Proposition~4.1 of Chapter~II in~\cite{KN}, the corresponding element of the holonomy group of $\Omega$ with reference point $(x,e)$ is obtained from~\eqref{6.7} by conjugation by $g^{-1}$, with $g$ given in~\eqref{6.0*}, and it is equal to the right-hand side of~\eqref{6.6}
 because
 \begin{equation}
\mathbf j(x) =\mathrm g^{-1}(-i\sigma_3)\mathrm g=\mathrm g^{-1}\mathbf e_3 \mathrm g,
\notag
\end{equation}
which incidentally clarifies the  sense of   the   rotationally invariant imaginary unit quaternion  $\mathbf j(x)$. The theorem is proved.

{\large Remark~1.} In Ref.~\cite{S}, we used a right multiplier $\mathsf M_{\rm R}$ of the  weak projective representation $a\to \mathsf V(a)= e^{a\cdot \boldsymbol{\nabla}}$ for comparison with works~\cite{EJ,CG-BLMV2}. It is  connected with  $\mathsf M$ in~\eqref{6.5} by $\mathsf M_{\rm R}(a,b)=\mathsf M(-b,-a)^\dagger$, as is easily seen by  Hermitian conjugation.  The formula~\eqref{6.6} was  given without proof by Emch and Jadczyk in~\cite{EJ} and was associated there with the right multiplier of the  representation $a\to e^{-a\cdot \boldsymbol{\nabla}}$. However, in that case the triangle $\bigtriangleup(x;a,b)$ must be replaced with  $\bigtriangleup(x;b,a)$, and this correction is essential for  accurate calculations.

{\large Remark~2.} In Ref.~\cite{CG-BLMV2}, an attempt was made to generalize the Emch and Jadczyk construction and to consider the covariant derivatives $\boldsymbol{\nabla}^{(\mathit g)}_k=\partial_k+\frac12\mathit g\epsilon_{ijk}\frac{x^i}{|x|^2}\mathbf e_j$ for arbitrary $\mathit g$ identified with the product of the electric and magnetic charges. However, it is easy to verify that the commutation relations $[\boldsymbol{\nabla}^{(\mathit g)}_i,\boldsymbol{\nabla}^{(\mathit g)}_j]=-\frac12 {\mathit g} \epsilon_{ijk}\frac{x^k}{|x|^3}\mathbf j(x)$ (numbered by (3.8) in~\cite{CG-BLMV2}) hold only for $\mathit g=1$, and only then correspond to the Poisson brackets~\eqref{2.2}. The essence of the matter is that the $SU(2)$-connection defining $\boldsymbol{\nabla}^{(\mathit g)}$ is reducible to $U(1)$ only if $\mathit g=1$. For $\mathit g\ne 1$, this  connection  is irreducible, except for the case $\mathit g=2$, when it  is reducible to the identity subgroup, i.e., is a pure gauge and  $[\boldsymbol{\nabla}^{(2)}_i,\boldsymbol{\nabla}^{(2)}_j]=0$. Indeed, it is easy to see that
\begin{equation}
-i\epsilon_{ijk}\frac{x^i}{|x|^2}\sigma^jdx^k=\mathrm g^{-1}d\mathrm g,\quad\text{where}\quad \mathrm g(\theta,\phi)= \begin{pmatrix}e^{i\phi}\sin\theta& -\cos\theta\\
 \cos\theta& e^{-i\phi}\sin\theta\end{pmatrix}.
   \notag
\end{equation}

We now let  $\mathsf J$ denote  the operator of multiplication by $\mathbf j(x)$,
\begin{equation}
 \big(\mathsf J\Psi\big)(x)\eqdef\mathbf j(x)\mathsf\Psi(x),
 \notag
 \end{equation}
 and introduce an operator-valued Fourier transform of  phase-space functions by substituting $\mathsf J$ for the complex imaginary unit $i$ in the usual Fourier transform,
 \begin{equation}
\tilde{\mathsf f}(u,v) =\frac{1}{(2\pi)^3}\int\! dqdp \big(\Re f(q,p)+\mathsf J\Im f(q,p)\big)\, e^{-\mathsf J(u\cdot p+v\cdot q)}.
 \notag
\end{equation}
The foregoing makes it possible to define a quaternionic Weyl quantization map by
  \begin{equation}
f \longmapsto  \mathsf O(f)=\frac{1}{(2\pi)^3}\int\! du dv\,\tilde{\mathsf f}(u,v)\,e^{\mathsf J(u\cdot  \mathsf P+v\cdot\mathsf  Q)},\qquad\text{where}\quad \mathsf P=-\mathsf J\hbar\boldsymbol{\nabla}
 \label{6.8}
\end{equation}
and where $\mathsf  Q$  is the position operator,   $(\mathsf  Q^i\mathsf\Psi)(x)=x^i\mathsf\Psi(x)$.

As an example, it is easy to verify that under the condition $eg=\hbar/2$  the  angular momentum components $\epsilon_{ijk}q^j p_k- eg\,q^i/|q|$ are mapped by~\eqref{6.8} to the Hermitian operators
 \begin{equation}
\mathsf L_i=-\mathsf J \hbar\left(\epsilon_{ijk}\mathsf Q^j\partial_k-\frac{1}{2}\mathbf e_i\right).
 \notag
\end{equation}
For a more detailed description of the quaternionic Weyl correspondence, we refer the reader to~\cite{S}, where explicit formulas for the phase-space star product induced by the correspondence~\eqref{6.8} are derived. In the next section, we prove
 that the quantization map~\eqref{5.4} yields exactly the same star product. The quaternionic analog of  the quantization~\eqref{5.8} is defined by
  \begin{equation}
f \longmapsto  \mathsf O_\alpha(f)=\frac{1}{(2\pi)^3}\int\! du dv\,\tilde{\mathsf f}(u,v)\,e^{\mathsf J(u\cdot  \mathsf P+v\cdot\mathsf  Q)}e^{\mathsf J \hbar ( v\cdot\alpha u-v\cdot u/2)}.
 \label{6.9}
\end{equation}

\section{\large From the operator product to the star product}

The simplest way  to find   the operation on the set of symbols that corresponds to operator multiplication, i.e., to find the phase-space star product induced by the map~\eqref{5.4} or~\eqref{5.8}  is to generalize the reasoning used by  von Neumann~\cite{N} in the case of usual Weyl correspondence. Using the relation~\eqref{5.2}, the product of the operators corresponding to   phase-space functions $f$ and $g$  can be written as an integral involving a bilinear combination of their Fourier transforms:
\begin{align}
 \mathcal O_\alpha(f)\mathcal O_\alpha(g)&=\frac{1}{(2\pi)^6}\int\!dwdw'\,\tilde f(w)\tilde g(w')  T_\alpha(w)T_\alpha(w')\notag\\
&=\frac{1}{(2\pi)^6}\int\!dwdw'\,\tilde f(w)\tilde g(w') \mathcal M_{\alpha,\hbar}(Q;w,w') T_\alpha(w+w')\notag\\
&=\frac{1}{(2\pi)^3}\int\! dw\,\left\{\frac{1}{(2\pi)^3}\int \! dw'\,\tilde f(w-w')\tilde g(w') \mathcal M_{\alpha,\hbar} (Q;w-w',w')\right\} T_\alpha(w).
\label{7.1}
\end{align}
In the absence of a magnetic field and for $\alpha=\frac12I$, the  expression in braces in~\eqref{7.1} is the  twisted convolution of $\tilde f$ and $\tilde g$ and the inverse Fourier transform converts it into the Weyl-Moyal star product. But in our case, this  expression is  not a scalar function   but an operator-valued function because so is the magnetic multiplier $\mathcal M_{\alpha,\hbar}$. This difficulty can  be overcome by using a simple lemma whose proof is based on the commutation relation between the magnetic translation operator $V(\hbar u)=e^{iu\cdot P}$ and a function of the position operator.

{\bf Lemma~2.} {\it Let $\mu(Q)$   be the  operator of multiplication by a complex-valued function  $\mu(x)$.    Then the symbol of the  operator $\mu(Q) T_\alpha(u,v)$  under the correspondence~\eqref{5.8} is equal to
\begin{equation}
\mu(x-\hbar\alpha u)e^{i(u\cdot p+v\cdot x)}.
\label{7.2}
\end{equation}

Proof.} The Fourier transform of the function~\eqref{7.2} is given by
\begin{equation}
 \frac{1}{(2\pi)^3}\int \!dpdx \,\mu(x-\hbar\alpha u)\, e^{i(u-u')\cdot p+i(v-v')\cdot x}=(2\pi)^{3/2}\delta(u'-u)\tilde\mu(v'-v)e^{i\hbar (v-v')\cdot \alpha u}.
\notag
\end{equation}
Hence, its corresponding operator is
  \begin{multline}
\frac{1}{(2\pi)^{3/2}}\int\! du' dv'\,\delta(u'-u)\tilde\mu(v'-v)e^{i\hbar (v-v')\cdot \alpha u}\,e^{iu'\cdot  P}e^{iv'\cdot  Q} e^{i\hbar v'\cdot(\alpha-I)u'}\,=\\=
e^{i\hbar v\cdot\alpha u}e^{iu\cdot P}\frac{1}{(2\pi)^{3/2}}\int dv'\,\tilde\mu(v'-v)\,e^{iv'\cdot (Q-\hbar u)}.
 \notag
\end{multline}
For any  $\Psi\in\mathcal H_n$,  we have
\begin{multline}
\frac{1}{(2\pi)^{3/2}}\int dv'\,\tilde\mu(v'-v)\,\left(e^{iv'\cdot (Q-\hbar u)}\Psi\right)(x)= \mu(x-\hbar u)
e^{iv\cdot (x-\hbar u)}\Psi(x)=\\= \left(\mu(Q-\hbar u)e^{iv\cdot (Q-\hbar u)}\Psi\right)(x).
\notag
\end{multline}
Because $e^{iu\cdot P}\mu(Q-\hbar u)=\mu(Q)e^{iu\cdot P}$, we  conclude that the function~\eqref{7.2} is  transformed by~\eqref{5.8} into the operator  $\mathcal \mu(Q)e^{iu\cdot  P}e^{iv\cdot  Q} e^{i\hbar v\cdot(\alpha-I)u}=\mu(Q)T_\alpha(u,v)$, which completes the proof.

An  analog of this lemma holds for the quaternionic quantization map~\eqref{6.9}, with $\mu(Q)$ replaced by the operator of multiplication by $\Re \mu(x)+\mathbf j(x)\Im \mu(x)$, for details see~\cite{S}, where it is proved for  $\alpha=\frac12I$.

Using the linearity of the quantization maps~\eqref{5.8} and~\eqref{6.9} and applying this lemma, or its quaternionic analog, with the function
 \begin{equation}
 \mu_w(x)=\frac{1}{(2\pi)^3}\int \! dw'\,\tilde f(w-w')\tilde g(w') \mathcal M_{\alpha,\hbar} (x;w-w',w')
  \notag
\end{equation}
depending parametrically on $w=(u,v)$,  we deduce that in both cases the star product of $f$ and $g$  can be be represented as an integral involving the shifted multiplier,
\begin{multline}
(f\star_\alpha g)(p,x)=\frac{1}{(2\pi)^6}\int dwdw'\,\tilde f(w-w')\tilde g(w')\mathcal M_{\alpha,\hbar}(x-\hbar\alpha u,w-w',w') e^{i(u\cdot p+v\cdot x)}\\=
\frac{1}{(2\pi)^6}\int dwdw'\,\tilde f(w)\tilde g(w')\mathcal M_{\alpha,\hbar}(x-\hbar\alpha(u+ u'),w,w') e^{i(u+u')\cdot p+i(v+v')x}.
 \label{7.3}
\end{multline}
By~\eqref{5.7}, the multiplier $\mathcal  M_{\alpha,\hbar}$ has a factored structure. This  allows us to perform integration with respect to $v$ and $v'$ and rewrite~\eqref{7.3}  in terms of the functions $f$ and $g$ themselves. Using the equality
 \begin{multline}
  \int dvdv'\,\tilde f(u,v)\tilde g(u',v')e^{i(v+v')x+i\hbar v'\cdot(I-\alpha)u-i\hbar  v\cdot \alpha u'}=\\=
\int dp'dp''f(p',x-\hbar\alpha u') g(p'',x+\hbar(I-\alpha) u)e^{-iup'-iu'p''},
 \notag
\end{multline}
we find that
\begin{multline}
(f\star_\alpha g)(p,x)=
 \frac{1}{(2\pi)^6}\int dudu'dp'dp''f(p',x-\hbar\alpha u') g(p'',x+\hbar(I-\alpha) u)\\
 \times e^{iu\cdot(p-p')+iu'\cdot (p-p'')}m(x-\hbar\alpha (u+u');\hbar u,\hbar u')
 \notag
\end{multline}
with $m(x;\cdot,\cdot)$ defined by~\eqref{4.3}. Changing the integration variables from $u$ and $u'$ to $x'=x-\hbar\alpha u'$ and $x''=x+\hbar(I-\alpha) u$ and using~\eqref{4.3}, we arrive at the following result.

{\bf Theorem~3.} {\it The quantization maps~\eqref{5.8} and~\eqref{6.9} induce the same phase-space star product. If $\det(\alpha(I-\alpha))\ne0$, the integral kernel
 ${\mK}_\alpha(x,p;x',p',x'',p'')$  of this product can be written as ${\mK}_\alpha=K_\alpha K_\alpha^{\rm magn}$,
where
\begin{multline}
 K_\alpha(x,p;x',p',x'',p'')=\frac{1}{(2\pi\hbar)^6|\det(\alpha(I-\alpha))|}
 \exp\Bigl\{\frac{i}{\hbar}\left[(p-p'')\cdot\alpha^{-1}(x-x')\right.\\-
 \left.(p-p')(I-\alpha)^{-1}(x-x'')\right]\Bigr\},
  \end{multline}
and the magnetic part is given by
\begin{multline}
K_\alpha^{\rm magn}(x;x',x'')=m\left(x'-\alpha(I-\alpha)^{-1}(x''-x);(I-\alpha)^{-1}(x''-x),\alpha^{-1}(x-x')\right)\\
=\exp\left\{-\frac{i}{\hbar}\int_{\bar\triangle_\alpha(x;x',x'')}\beta\right\},
 \label{7.4}
\end{multline}
where $\bar\triangle_\alpha(x;x',x'')$ is the triangle whose vertices $\bar x$, $\bar x'$, $\bar x''$ are related to the points $x$, $x'$, $x''$ by}
 \begin{equation}
x=\alpha \bar x''+(I-\alpha)\bar x,\quad x'=\alpha \bar x'+(I-\alpha)\bar x,\quad \bar x''=\alpha \bar x''+(I-\alpha)\bar x'.
 \notag
\end{equation}

In the case of  Weyl quantization~\eqref{5.4},  the triangle $\bar\triangle_{\frac12I}(x;x',x'')$
has  $x$, $x'$, and $x''$ as midpoints of its sides.
This result  agrees with those obtained  in Refs.~\cite{M,KO1,KO2,MP1} for the case of a divergence-free magnetic field and trivial phase-space topology. From what has been said in Sec.~4, it is clear that  the associativity of the star product $\star_\alpha$ for the charge-monopole system is ensured by the charge quantization condition~\eqref{3.2}.

If the matrix $\alpha$ (or $(I-\alpha)$) is singular, then its corresponding kernel is not locally integrable, but can be viewed as a tempered distribution. In particular, if $\alpha=0$, then
\begin{multline}
\mK_0(x,p;x',p',x'',p'')=\frac{1}{(2\pi)^6\hbar^3}\delta(x-x')
 \exp\Bigl\{-\frac{i}{\hbar}(p-p')\cdot(x-x'')\Bigr\}\\
\times\int du  \exp\left\{iu\cdot(p-p'')-\frac{i}{\hbar}\int_{\triangle(x;x''-x,\hbar u)}\beta\right\},
 \notag
 \end{multline}
 and for $\alpha=I$, we get
 \begin{multline}
\mK_I(x,p;x',p',x'',p'')=\frac{1}{(2\pi)^6\hbar^3}\delta(x-x'')
 \exp\Bigl\{\frac{i}{\hbar}(p-p'')\cdot(x-x')\Bigr\}\\
\times\int du  \exp\left\{iu\cdot(p-p')-\frac{i}{\hbar}\int_{\triangle(x'-\hbar u;\hbar u,x-x')}\beta\right\}.
 \notag
 \end{multline}

\section{\large The asymptotic expansion of the  magnetic product}

The asymptotic differential form of the star product $\star_\alpha$ can be derived by expanding the shifted multiplier in the right-hand side of~\eqref{7.3} in powers of the Planck constant. The coefficients of this expansion are polynomials in the variables $u,v,u',v'$ and the inverse Fourier transform given by~\eqref{7.3} converts them into differential operators acting on $f$ and $g$. In particular,
\begin{equation}
u\to -i\overleftarrow{\partial_p},\quad v\to -i \overleftarrow{\partial_x},\quad u'\to -i\overrightarrow{\partial_p},\quad v'\to -i\overrightarrow{\partial_x},
  \label{8.1}
\end{equation}
where $\overleftarrow{\partial}$ acts on $f$ and  $\overrightarrow{\partial}$ acts on $g$.
This yields the desired representation
 \begin{equation}
f\star_\alpha g= \sum_{n=0}^\infty \hbar^n \mB_n(f,g)
 \label{8.2}
\end{equation}
with some bidifferential operators $\mB_n$.

According to~\eqref{4.2}, \eqref{4.3} and~\eqref{5.7}, the magnetic part of $\mathcal M_{\alpha,\hbar}(x-\hbar\alpha(u+ u'),w,w')$ is
\begin{equation}
\exp\left\{-i\hbar\int_0^1dt_1\int_0^{t_1}dt_2\, u^i\beta_{ij}\left(x-\hbar\alpha(u+u')+\hbar t_1 u+\hbar t_2 u'\right){u'}^j\right\}.
 \label{8.3}
\end{equation}
The coefficients of the Taylor expansion of the exponent in~\eqref{8.3} around the point $x$ were evaluated in~\cite{M,KO1}  for the case of the magnetic Weyl-Moyal product, i.e., for $\alpha=\frac12 I$. To find  explicitly the operators $\mB_n$, the exponential should also be expanded and, unfortunately, even in that case, the general term of~\eqref{8.2} cannot be written in a closed compact form. Here we
 compute the star product up to the third order in $\hbar$ for $\alpha=tI$ with arbitrary real $t$. Using the dot notation for  summation over the suppressed  indices,
we have
\begin{equation}
\beta(x+\hbar y)=\beta(x)+\hbar (y\cdot\partial) \beta\left|_x\right. +\frac{\hbar^2}{2!} (y\cdot \partial)^2\beta\left|_x\right.+O(\hbar^3).
\label{8.4}
\end{equation}
 Substituting~\eqref{8.4}  with $y=(t_1-t) u+(t_2-t) u'$ into~\eqref{8.3} and integrating with respect to $t_1$ and $t_2$, we find that
\begin{multline}
\mathcal M_{t,\hbar}(x-\hbar t(u+ u'),w,w')=\exp\left\{i\hbar\left[(1-t)u\cdot v'- t v\cdot u'-\frac12 u\cdot\beta u'\right]\right.\\
 -\frac{i\hbar^2}{2}\left[\left(\frac23- t\right)u\cdot(u\cdot\partial)\beta u'+ \left(\frac13- t\right) u\cdot(u'\cdot\partial)\beta u'\right] \\
-\frac{i\hbar^3}{4}\left[\left(\frac12-\frac43 t+t^2\right)u\cdot(u\cdot\partial)^2\beta u'+ \left(\frac16-\frac23 t+t^2\right)u\cdot(u'\cdot\partial)^2\beta u'\right.\\
+\left.\left.2\left(\frac12-t\right)^2 u\cdot(u\cdot\partial)(u'\cdot\partial)\beta u'\right] +O(\hbar^4)\right\}.
 \label{8.4*}
\end{multline}
Expanding the exponentials and making the replacements~\eqref{8.1}, we obtain the following result:

{\bf Theorem~4.} {\it If $\alpha=tI$, then the star product induced by the quantization map~\eqref{5.8} is given, up to the third order in $\hbar$,   by the expression
\begin{multline}
\sum_{k=0}^3\frac{(i\hbar)^k}{k!}\Bigl[t\,\overleftarrow{\partial_x}\cdot \overrightarrow{\partial_p}- (1-t)
\overleftarrow{\partial_p}\cdot\overrightarrow{\partial_x}
+\frac12 \overleftarrow{\partial_p}
\cdot \beta\overrightarrow{\partial_p}\Bigr]^k\\
+\frac{\hbar^2}{2}\left[\left(\frac23- t\right)
\overleftarrow{\partial_p}\cdot(\overleftarrow{\partial_p}\cdot\partial_x)
\beta\overrightarrow{\partial_p}+\left(\frac13- t\right)
\overleftarrow{\partial_p}\cdot(\overrightarrow{\partial_p}\cdot\partial_x)
\beta\overrightarrow{\partial_p}\right]\\
\times\left[1+i\hbar\left(t\,
\overleftarrow{\partial_x}\cdot \overrightarrow{\partial_p}-(1-t)
\overleftarrow{\partial_p}\cdot\overrightarrow{\partial_x}
+\frac12 \overleftarrow{\partial_p}
\cdot \beta\overrightarrow{\partial_p}\right) \right]
\\
-\frac{i\hbar^3}{4}\left[\left(\frac12-\frac43 t+t^2\right) \overleftarrow{\partial_p}\cdot(\overleftarrow{\partial_p}\cdot\partial_x)^2
\beta\overrightarrow{\partial_p}+\left(\frac16-\frac23 t+t^2\right)
\overleftarrow{\partial_p}\cdot(\overrightarrow{\partial_p}\cdot\partial_x)^2
\beta\overrightarrow{\partial_p}\right.\\ +\left. 2\left(\frac12-t\right)^2 \overleftarrow{\partial_p}\cdot(\overleftarrow{\partial_p}\cdot\partial_x)(\overrightarrow{\partial_p}\cdot\partial_x)
\beta\overrightarrow{\partial_p} \right],
 \label{8.5}
\end{multline}
where  $\partial_x$ without an  over-arrow is applied  to  the matrix  $\beta(x)$.}

An alternate way of calculating the asymptotic expansion of $\star_\alpha$ is by using the Zassenhaus formula for the product of exponentials of noncommuting variables, see  the appendix.

The case of the magnetic Weyl-Moyal product  deserves special attention.
 If $t=1/2$, then the sum in square brackets in the first line of~\eqref{8.5} can obviously be written as $\overleftarrow{\partial_a}\mathcal P^{ab}\overrightarrow{\partial_b}$, where  $\mathcal P^{ab}(x)$ is the $(a,b)$-th entry of the Poisson matrix~\eqref{2.4} and
 \begin{equation}
  \partial_a=\begin{cases}
  \partial/\partial x_a &\text{for $1\le a\le 3$},\\
   \partial/\partial p_a &\text{for $4\le a\le 6$}.
 \end{cases}
 \notag
 \end{equation}
It is important that in this case the  star product as a whole can be expressed explicitly and purely  in  terms of the initial Poisson structure. Using the expression~\eqref{8.5} and  taking the  block form of $\mathcal P^{ab}$ into account, we obtain
\begin{equation}
\begin{split}
f\star g &=fg+\frac{i\hbar}{2} \mathcal P^{ab}\partial_a f
\partial_b g
-\frac{\hbar^2}{8}\mathcal P^{a_1b_1}\mathcal P^{a_2b_2}
\partial_{a_1}\partial_{a_2}f
\partial_{b_1}
\partial_{b_2}g\\
 &-\frac{i\hbar^3}{48}\mathcal P^{a_1b_1}\mathcal P^{a_2b_2}\mathcal P^{a_3b_3}
\partial_{a_1}\partial_{a_2} \partial_{a_3}f
\partial_{b_1}
\partial_{b_2}\partial_{b_3} g\\
&-\frac{\hbar^2}{12}\mathcal P^{a_1b_1} \partial_{b_1}
\mathcal P^{a_2b_2}\bigl(
\partial_{a_1}\partial_{a_2}f\partial_{b_2} g -
\partial_{a_2}f\partial_{a_1}\partial_{b_2}g\bigr) \\
&-\frac{i\hbar^3}{24}\mathcal P^{a_1b_1}
\mathcal P^{a_2b_2} \partial_{b_2} \mathcal P^{a_3b_3}\bigl(
\partial_{a_1}\partial_{a_2}\partial_{a_3}f\partial_{b_1}\partial_{b_3} g -
\partial_{a_1}\partial_{a_3}f\partial_{b_1}\partial_{a_2}\partial_{b_3}g\bigr)\\
&-\frac{i\hbar^3}{48}\mathcal P^{a_1b_1} \mathcal P^{a_2b_2}\partial_{b_1}\partial_{b_2}
\mathcal P^{a_3b_3}\bigl(
\partial_{a_1}\partial_{a_2}\partial_{a_3}f\partial_{b_3} g +
\partial_{a_3}f\partial_{a_1}\partial_{a_2}\partial_{b_3}g\bigr)
 +O(\hbar^4).
 \end{split}
 \label{8.6}
\end{equation}
The resulting expression  is in complete agreement with the  Kontsevich  formula~\cite{K} for deformation quantization of general  Poisson manifolds and  can also be described in terms of graphs introduced by Kontsevich, namely,
\begin{gather}
\stackrel{f}{\ssb}\,\star\,\stackrel{g}{\ssb}\;=\;
 \stackrel{f}{\ssb}\times\stackrel{g}{\ssb}\, + \; \frac{i\hbar}{2}\;
 \xymatrix@=0.4cm{
 \stackrel{f}{\ssb} &\bullet\ar[r] \ar[l]& \stackrel{g}{\ssb}
 }
 + \frac{1}{2!}\left(\frac{i\hbar}{2}\right)^2 \xymatrix@=0.4cm{
\stackrel{f}{\ssb} &\bullet\ar[r] \ar[l]& \stackrel{g}{\ssb}\\
  &\bullet\ar[ur]\ar[ul]
  }
  +\frac{1}{3!}\left(\frac{i\hbar}{2}\right)^3\xymatrix@=0.4cm{
\stackrel{f}{\ssb} &\bullet\ar[r] \ar[l]& \stackrel{g}{\ssb}\\
  &\bullet\ar[ur]\ar[ul]\\
  &\bullet\ar[uul]\ar[uur]
  }
 \notag\\
  +\frac{1}{3}\left(\frac{i\hbar}{2}\right)^2 \Biggl(\xymatrix@=0.4cm{
\stackrel{f}{\ssb} &\bullet\ar[r] \ar[l]& \stackrel{g}{\ssb}\\
  &\bullet\ar[u]\ar[ul]
  } +
  \xymatrix@=0.4cm{
\stackrel{f}{\ssb} &\bullet\ar[r] \ar[l]& \stackrel{g}{\ssb}\\
  &\bullet\ar[u]\ar[ur]
  } \Biggr)
  \label{8.7}  \\
   + \frac{1}{3}\left(\frac{i\hbar}{2}\right)^3 \Bigg(\xymatrix@=0.4cm{
\stackrel{f}{\ssb} &\bullet\ar[r] \ar[l]& \stackrel{g}{\ssb}\\
  &\bullet\ar[u]\ar[ul]\\
  &\bullet\ar[uul]\ar[uur]
  }
   +\xymatrix@=0.4cm{
\stackrel{f}{\ssb} &\bullet\ar[r] \ar[l]& \stackrel{g}{\ssb}\\
  &\bullet\ar[u]\ar[ur]\\
  &\bullet\ar[uul]\ar[uur]
  } \Biggr)
  + \frac{1}{6}\left(\frac{i\hbar}{2}\right)^3 \Bigg(\xymatrix@=0.4cm{
\stackrel{f}{\ssb} &\bullet\ar[r] \ar[l]& \stackrel{g}{\ssb}\\
  &\bullet\ar[u]\ar[ul]\\
  &\bullet\ar@/_1pc/[uu]\ar[uul]
  }
  + \xymatrix@=0.4cm{
\stackrel{f}{\ssb} &\bullet\ar[r] \ar[l]& \stackrel{g}{\ssb}\\
  &\bullet\ar[u]\ar[ur]\\
  &\bullet\ar@/^1pc/[uu]\ar[uur]
  } \Biggr)
\notag
\end{gather}
The internal vertices of these graphs contain the Poisson matrix $\mathcal P^{ab}$ and the directed edges symbolize the derivatives $\partial_a$ and $\partial_b$   acting on the content of the vertex at the arrowhead. The summation over $a$ and $b$ is implicit and the ordering of indices in $\mathcal P^{ab}$ corresponds to the  left-right ordering of the outgoing edges. The weights  of these graphs coincide exactly with those defined by the Kontsevich integral formula, but it should be noted that the deformation parameter denoted by   $\hbar$ in~\cite{K} corresponds to $i\hbar/2$ in our notation. Some generally possible graphs with nonzero weights  are absent in~\eqref{8.7} because their associated operators vanish in the case of a magnetic Poisson structure. For instance, the second-order loop graph
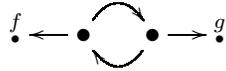
\begin{equation}
\xymatrix@=0.5cm{
 \stackrel{f}{\ssb} &\bullet\ar[l]\ar@/^1pc/[r]&\bullet\ar[r]\ar@/^1pc/[l] &\stackrel{g}{\ssb}
 }
\notag
\end{equation}
corresponds to the operator $\overleftarrow{\partial_{a_1}}\partial_{a_2}\mathcal P^{a_1b_1} \partial_{b_1} \mathcal P^{a_2b_2} \overrightarrow{\partial_{b_2}}$ which is zero in our case, because the Poisson matrix~\eqref{2.4} depends on $x$ but not on $p$ and  hence $\partial_{a_2}\mathcal P^{a_1b_1}=0$ for $a_2>3$, whereas if $a_2\le3$, then $\mathcal P^{a_2b_2}$ is constant and $\partial_{b_1} \mathcal P^{a_2b_2}=0$. Clearly, the third-order
graphs with loops also make zero contribution. A similar argument applies to the graphs
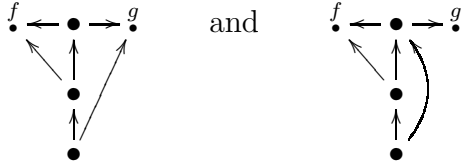
\begin{equation}
\xymatrix@=0.4cm{\stackrel{f}{\ssb} &\bullet\ar[r] \ar[l]& \stackrel{g}{\ssb}\\
  &\bullet\ar[u]\ar[ul]\\
  &\bullet\ar[u]\ar[uur]
 }\qquad \text{and}\qquad
 \xymatrix@=0.4cm{
\stackrel{f}{\ssb} &\bullet\ar[r] \ar[l]& \stackrel{g}{\ssb}\\
  &\bullet\ar[u]\ar[ul]\\
  &\bullet\ar@/_1pc/[uu]\ar[u]
  }
\notag
\end{equation}
which also do not contribute to the magnetic Weyl-Moyal product.

\section{\large Summary and conclusions}

We have seen that the generalized Weyl quantization map as well as magnetic analogs of other quantizations can be naturally defined for the charged particle-monopole system by using  the parallel transport operator. This formulation is completely gauge-independent and
follows the basic principle of geometric quantization that the phase-space symplectic form divided by $\hbar$ should be identified with the  curvature form of  a connection on an appropriate line bundle. Although the quaternionic quantization scheme deals with a trivial bundle, the operator representation in a complex Hilbert space  is certainly preferable from the practical viewpoint  because of the noncommutativity of quaternion multiplication. The simplest way to find the phase-space star product induced by the magnetic Weyl correspondence is  to use the fact that the magnetic translation operators form a weak projective representation. The integral-kernel form of this product can be written in purely symplectic terms and its asymptotic  expansion is expressed in terms of the initial Poisson structure and  agrees completely with the Kontsevich formula for deformation quantization of Poisson manifolds. The associativity of the integral form of the magnetic product is ensured by the charge quantization condition, whereas the associativity of its differential form holds for any charges because it is understood in the sense of formal power series in $\hbar$.

In conclusion, we note that most of theorems of the magnetic Weyl calculus were established for the case of a linear phase space and under the assumption that the magnetic field is infinitely differentiable and all its derivatives are polynomially bounded. Then the Schwartz space of smooth rapidly decreasing functions is closed under the magnetic Weyl product and, as a consequence, this product can be extended by duality to  tempered distributions belonging to the so-called magnetic Moyal algebra. This construction is not directly applicable to the  monopole field which is singular at  the origin, but the  analysis carried out above provides a basis for identifying the part of this calculus that admits an extension to the monopole case.

\section*{\large Appendix. Calculation of the star product with the use of the Zassenhaus formula}

 The Zassenhaus formula  is the dual of the Baker-Campbell-Hausdorff formula and gives decomposition of the exponential of the sum of two noncommuting operators $X$ and $Y$ into a product of exponential operators.
  It states that
 $$
 e^{X+Y}= e^{X}e^{Y}\prod_{n=2}^\infty e^{C_n(X,Y)},
\eqno (\rm A.1)
$$
where $C(X,Y)$ is a homogeneous Lie polynomial in $X$ and $Y$ of degree $n$. The first terms in (A.1) are written as
\begin{equation}
 C_2=-\frac12 [X,Y],\quad C_3=\frac13 [Y,[X,Y]]+\frac16 [X,[X,Y]],
\notag
\end{equation}
\begin{equation}
 C_4=-\frac18 [Y,[Y,[X,Y]]]-\frac{1}{24} [X,X,[X,Y]]]-\frac18[Y,[X,[X,Y]]].
\notag
\end{equation}
Some systematic approaches to  computing $C_n$ for $n>4$ are presented, e.g., in~\cite{CMN, TK}.
Setting $X=iu\cdot P$ and $Y=iu'\cdot P$, we see that the Zassenhaus formula (A.1) provides a means of calculating the right multiplier of the weak projective representation $u\to e^{iu\cdot P}$.  As  was mentioned in Sec.~6, a right multiplier was considered in~\cite{S} in the quaternionic setting. The left multiplier in equation~\eqref{5.2} is expressed by the ``left-oriented'' Zassenhaus formula
$$
 e^{X+Y}=\dots e^{C'_4(X,Y)} e^{C'_3(X,Y)} e^{C'_2(X,Y)} e^{X}e^{Y}.
\eqno (\rm A.2)
$$
The terms $C_n$ in (A.1) and $C'_n$ in (A.2) are connected by the simple relation
\begin{equation}
 C'_n(X,Y)= (-1)^{n+1}C_n(Y,X).
\notag
\end{equation}
In our case,
\begin{equation}
   C'_2=C_2=-\frac12 [iu\cdot P,iu'\cdot P]=\frac{i\hbar}{2}u\cdot\beta u'
\notag
\end{equation}
and the calculation of the higher-order nested commutators reduces to differentiation of $\beta(x)$. Using explicit expressions for $C'_3(iu\cdot P,iu'\cdot P)$ and for  $C'_4(iu\cdot P,iu'\cdot P)$, we immediately obtain
$$
\hspace{-1.6cm} m(x,\hbar u,\hbar u')=\exp\left\{-\frac{i\hbar}{2} u\cdot\beta u'-\frac{i\hbar^2}{3}\left[ u\cdot(u\cdot\partial)\beta u'+ \frac12 u\cdot(u'\cdot\partial)\beta u' \right]\right.
$$
$$
\hspace{1cm} -\frac{i\hbar^3}{8}\left. \left[u\cdot(u\cdot\partial)^2\beta u'+ \frac13u\cdot(u'\cdot\partial)^2\beta u'+  u\cdot(u\cdot\partial)(u'\cdot\partial)\beta u'\right]+O(\hbar^4)\right\}.
\eqno (\rm A.3)
$$
Replacing $x$ in (A.3) by $x-\hbar t(u+u')$ and then using~\eqref{8.4} with $y=-t(u+u')$, we again arrive at~\eqref{8.4*}.

\end{document}